\documentclass[
reprint,
groupedaddress,
showpacs,
showkeys,
amsmath, amssymb,
aps,
prx,
]{revtex4-2}

\usepackage{xcolor}
\usepackage{tabularx}
\usepackage{graphicx}
\usepackage{dcolumn}
\usepackage{bm}
\usepackage[colorlinks=true, allcolors=blue]{hyperref}
\usepackage{comment}
\usepackage[normalem]{ulem}
\graphicspath{{figures/}}
\usepackage[separate-uncertainty=false]{siunitx}
\usepackage{braket}
\usepackage{xspace}
\usepackage{xstring}

\usepackage{adjustbox}
\usepackage{multirow}

\newcommand{\down}{\ensuremath{\ket{\downarrow}}\xspace}
\newcommand{\up}{\ensuremath{\ket{\uparrow}}\xspace}

\newcommand{\blue}[1]{\textcolor{blue}{}}
\newcommand{\suppress}[1]{\textcolor{brown}{}}

\newcommand*{\aref}[1]{%
	\IfBeginWith{#1}{eq:}{Eq.~\eqref{#1}}{}
	\IfBeginWith{#1}{fig:}{Fig.~\ref{#1}}{}%
	\IfBeginWith{#1}{tab:}{Table~\ref{#1}}{}%
	\IfBeginWith{#1}{appendix:}{Appendix~\ref{#1}}{}%
	\IfBeginWith{#1}{sec:}{Section~\ref{#1}}{}%
	}

\usepackage{tikz}
\newcommand*\circled[1]{\tikz[baseline=(char.base)]{
    \node[shape=circle, draw, inner sep=1pt, 
        minimum height=12pt] (char) {#1};}}	
\newcommand\circledcaption[1]{\raisebox{.5pt}{\textcircled{\raisebox{-.9pt}{#1}}}} 

\makeatletter
\def\maketitle{
\@author@finish
\title@column\titleblock@produce
\suppressfloats[t]}
\makeatother
	
\begin{document}

\title{Ferromagnetism in an extended coherently-coupled atomic superfluid}

\date{\today}
\author{R. Cominotti}
\thanks{These two authors contributed equally to this work.}
\author{A. Berti}
\thanks{These two authors contributed equally to this work.}
\author{C. Dulin}
\thanks{Present address: ENS, Paris.}
\author{C. Rogora}
\author{G. Lamporesi}
\email[]{giacomo.lamporesi@ino.cnr.it}
\author{I. Carusotto}
\author{A. Recati}
\email[]{alessio.recati@ino.cnr.it}
\author{A. Zenesini}
\email[]{alessandro.zenesini@ino.cnr.it}
\author{G. Ferrari}

\affiliation{Pitaevskii BEC Center, CNR-INO and Dipartimento di Fisica, Universit\`a di Trento, I-38123 Trento, Italy.}

\begin{abstract}

Ferromagnetism is an iconic example of a first-order phase transition taking place in spatially extended systems and is characterized by hysteresis and the formation of domain walls. In this paper we demonstrate that an extended atomic superfluid in the presence of a coherent coupling between two internal states exhibits a quantum phase transition from a para- to a ferromagnetic state. The nature of the transition is experimentally assessed by looking at the phase diagram as a function of the control parameters, at hysteresis phenomena, at the magnetic susceptibility and the magnetization fluctuations around the critical point. We show that the observed features are in good agreement with mean-field calculations. Additionally, we develop experimental protocols to deterministically generate domain walls that separate spatial regions of opposite magnetization in the ferromagnetic state.  Thanks to the enhanced coherence properties of our atomic superfluid system compared to standard condensed matter systems, our results open the way towards the study of different aspects of the relaxation dynamics in isolated coherent many-body quantum systems.

\end{abstract}

\maketitle

\section{INTRODUCTION}

Superfluidity in many-body quantum systems leads to interesting and notable transport and coherence properties \cite{Pines1990,Svistunov2015,Dalibard2008}. Such properties are due to a thermal second-order transition from a normal to a superfluid state, a transition which is formally characterised by the spontaneous breaking of the $U(1)$ symmetry related to particle number conservation. Since such a transition is driven by the Bose statistics, atom-atom interactions are not needed to observe condensation. However, they play an important role in stabilizing the superfluid phase against disturbances, e.g. guaranteeing a finite compressibility and a finite critical Landau velocity for superfluidity. 

On top of this, superfluids can also have internal degrees of freedom, leading to order parameters with a non-trivial spinor or vector structure~\cite{Stamperkurn2013}. In this case, non-spin-symmetric interactions may lead to ground states with a very different spinor structure of the order parameter. A natural question is therefore whether the transition between different states can be described as a quantum phase transition (QPT), and, if so, which universality class such a QPT belongs to, what is the interplay between the superfluid nature of the system and the QPT, and whether the QPT in these systems survives at low yet finite temperature \cite{Vojta2003}.

Recent theoretical works~(see review in \cite{Recati22}) have anticipated that a two-component atomic Bose-Einstein condensate (BEC), subject to an external field that coherently couples the two components \cite{Matthews1999}, exhibits a phase transition in the quantum Ising universality class. In particular, at zero temperature, mean-field theory predicts an interaction-driven transition from a paramagnetic (PM) to a $\mathbb{Z}_2$-symmetry-breaking ferromagnetic (FM) state.

The quantum Ising model \cite{Ising1925,sachdev} is the paradigmatic model for (continuous) QPT, where the ferromagnetic interactions along one spin direction of the standard Ising model compete with a transverse magnetic field. 
The dynamics of the low-energy magnetic fluctuations near the critical point are described by a $\phi^4$ theory. This is based on the Ginzburg-Landau functional for a continuous phase transition, with its iconic single- to double-well energy landscape transition upon the change of an external parameter \cite{Landau1984}. 

\begin{figure*}
    \centering
    \includegraphics[width = \textwidth]{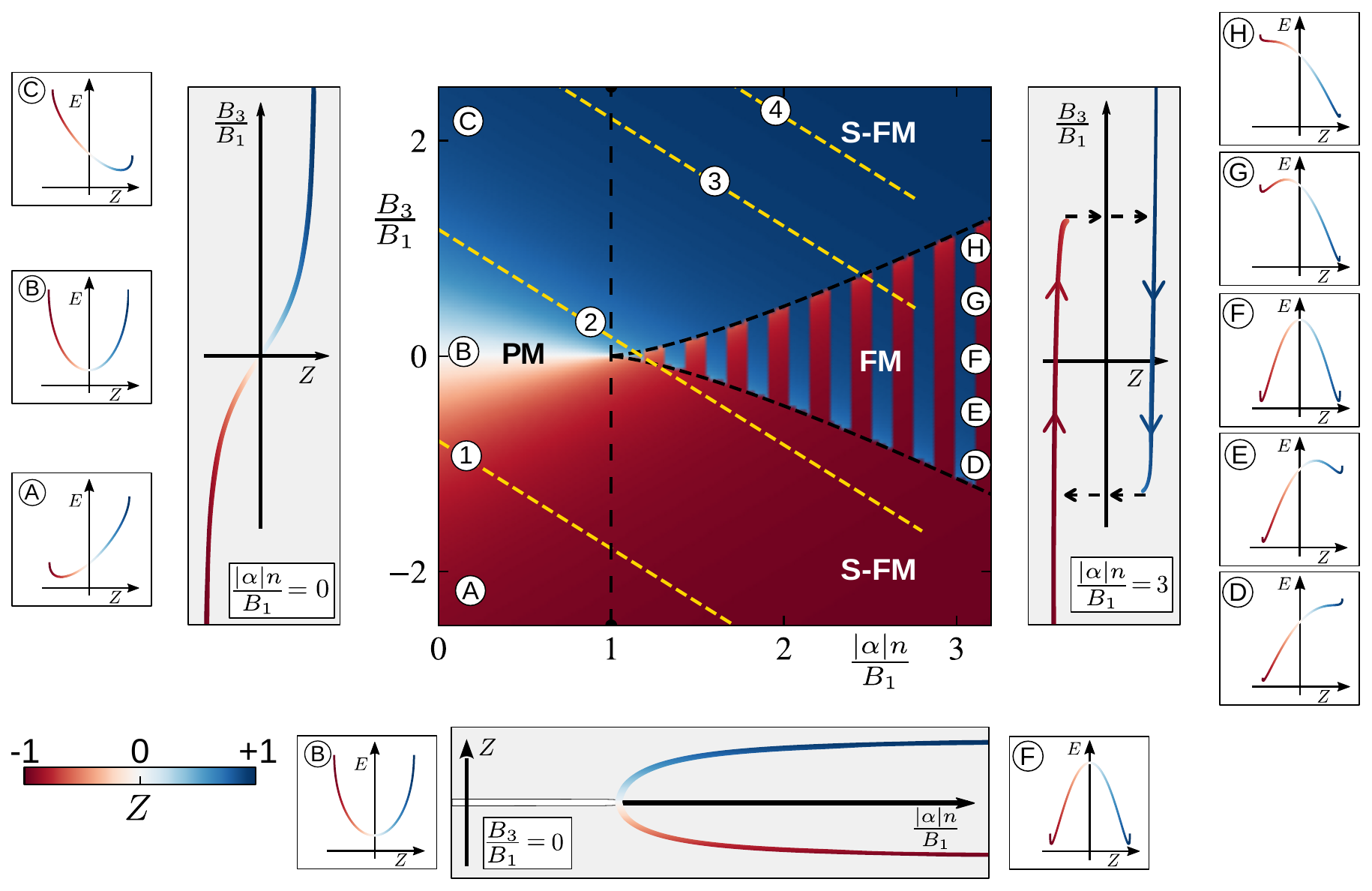}
    \caption{Phase diagram of the magnetic model. The relative magnetization $Z$ of the system's stationary states is shown as a function of the nonlinearity and of the axial magnetic field strength, both in units of the transverse field. The system can be para- ($|\alpha| n<B_1$), ferro- ($|\alpha| n > B_1 \gg B_3$) or saturated ferromagnetic ($|\alpha| n > B_1$ and $B_3\gg B_1$). Panels A-H show the dependence of the energy  [Eq.~(\ref{eq:energy1})] on the relative magnetization $Z$ in several points of the phase diagram. Three grey side panels show the value of $Z$ at the energy minimum, as a function of $|\alpha| n/B_1$ for $B_3=0$ (bottom) and as a function of $B_3/B_1$ for $|\alpha| n/B_1=0$ (left) or $|\alpha| n/B_1=3$  (right). Numbered dashed yellow lines mark four different single-shot experimental realizations in the atomic system as reported in Fig.~\ref{fig:fig2}. See Sec.~\ref{atom} and Table~\ref{table1} for mapping from magnetic to atomic system.}
    \label{fig:fig1}
\end{figure*} 

In this work, we experimentally demonstrate that a superfluid of coherently-coupled Sodium atoms exhibits such a ferromagnetic phase transition. While classical bifurcation \cite{Zibold10} and hysteresis phenomena \cite{Trenkwalder2016} were already observed in zero-dimensional atomic systems, the lack of a spatial extension inhibited the possibility of describing the observed phenomena in terms of a phase transition. This is, instead, possible in our spatially extended ultracold cloud of coherently-coupled atoms, where  the spin degree of freedom can be properly described as a ferromagnetic order parameter with anisotropic interactions and subject to an external field. Taking advantage of the coexistence of different magnetic phases within a single system, we map out the phase diagram of the ground state and characterize the associated hysteresis phenomena. Exploiting the long coherence time of our system and its robustness against localized spin excitations, we demonstrate the possibility to deterministically generate domain walls between different magnetic states.

Our results highlight the potential of a coherently-coupled BECs as a new platform, where to explore QPTs.
As compared to usual solid-state systems, our platform features important advantages. On the one hand, the cold atom platform allows for a microscopic description of all the interaction processes taking place in the system and, therefore, is amenable to a quantitative comparison with theory. On the other hand, the superfluid nature of the cloud and the smoothness of the trapping potential remove all those complications that normally stem from the unavoidable disorder of solid-state systems and their fast incoherent relaxation process, hence allowing to focus on the intrinsic many-body properties of the stationary  states. On a longer run, we expect that this feature will be of extreme importance, in view of applications to the experimental study of coherent relaxation phenomena in isolated quantum systems, such as the time-dependent motion of domain walls or the quantum-induced bubble-mediated decay of metastable states.

The structure of the article is the following: In Section II, we illustrate the properties of the PM-FM phase transition and define the relevant quantities in our atomic system. We describe our novel atomic platform and the experimental protocol in Section III. Section IV shows the experimental results, as well as a comparison with mean-field theory, and Section V is devoted to the controlled generation and observation of magnetic domain walls. Conclusions and future perspectives are reported in Section VI. 

\section{Para- to ferromagnetic phase transition}
\label{model}  
In order to show that our experimental platform can be described as a ferromagnet, in this Section we first briefly recall the semiclassical continuous description of a ferromagnetic system (Sec.~\ref{magm}) and then show how our atomic system naturally maps onto the magnetic model  (Sec.~\ref{atom}). A concise summary of the mapping is given in Tab.~\ref{table1}.

\subsection{The magnetic model}
\label{magm}

A textbook model of a FM to PM transition at zero temperature is based on a spin chain, subject to an external magnetic field and to internal spin-spin interactions. Within a mean-field approach, the energy of a ferromagnetic material can be written \cite{Abert2019} in terms of the local spin $\textbf{S}=(S_1,S_2,S_3)$ as
\begin{equation}
      E(\mathbf{S}) \propto - \int \left( \mathbf{B}\cdot\mathbf{S} - \frac {1}{2} \mathbf{S}\cdot \Bar{\Bar{K}}\mathbf{S} -\frac 1 2 |\nabla \mathbf{S}|^2 \right) dV.
      \label{eq:energy}
 \end{equation} 
 In the previous expression, $\mathbf{B}$ is the external field, $\Bar{\Bar{K}}$ is a diagonal matrix describing the anisotropic magnetic interactions in the material due to, e.g., the sample crystalline
structure, and the last term is the exchange energy, which accounts for the tendency of having a spatially uniform magnetization.
In the absence of any damping, the dynamics of the local spin is given by the (dissipationless) Landau-Lifshitz equation \cite{Landau1935} $ \partial_t \textbf{S}=-\textbf{H}_\textrm{eff} \times \textbf{S}$,  i.e., a non-linear precession around the effective field $\mathbf{H}_\mathrm{eff}=-\delta E/\delta\mathbf{S}$.

For the later analogy with our two-component superfluid platform, we consider a translationally invariant ferromagnet of spin density $n=|\mathbf{S}|$ with uniaxial magnetic anisotropy such that the only non-zero element of the magnetic interactions is $K_{33}=\alpha<0$, which sets the  {\it{easy}} axis along the axial direction 3. The magnetic field is uniform and has components along the axial  direction ($B_3$) and in the transverse plane ($B_1$).  The ground state solutions are characterized by homogeneous profiles, and a uniform effective magnetic field $\mathbf{H}_\mathrm{eff}=(B_1,0,B_3-\alpha  S_3)$. In this case, the energy of the system can be written as
\begin{equation}
      E(Z,\phi) \propto - B_3 Z - \frac {|\alpha| n}{2} Z^2 - B_1 \sqrt{1-Z^2}\cos \phi,
      \label{eq:energy1}
 \end{equation}
where $Z = S_3/n$ is the relative magnetization and $\phi=\arctan{(S_2/S_1)}$ is the angle of the spin in the plane.  
The ground state is obtained by minimizing the energy $E(Z,\phi)$ with respect to $Z$ and $\phi$. Assuming $B_1>0$, all ground states have $\phi=0$. The relative magnetization $Z$, instead, is a function of $B_3/B_1$ and $|\alpha| n/B_1$, as shown in \aref{fig:fig1}. The energy profiles computed using \aref{eq:energy1} are shown for eight different points in the phase diagram (see panels A-H). 
If $B_3=0$, the energy landscape shows a transition 
from a single minimum (paramagnet) with $Z=0$, when $|\alpha| n/B_1<1$, to a symmetric double minimum (ferromagnet) with $Z\neq 0$, when $|\alpha| n/B_1>1$, corresponding to the  $\mathbb{Z}_2$ symmetry breaking, $Z\leftrightarrow -Z$ (see bottom grey panel in \aref{fig:fig1}).

In the presence of a finite $B_3$, the energy minimum is shifted to a finite magnetization in the PM phase, 
while in the FM region, an energy splitting is observed between the two minima, corresponding to the absolute ground state and to a metastable state, whose lifetime is expected to depend on the height of the barrier between the two minima \cite{Lagnese21}.
For very strong $B_3$ beyond some critical value (panels D,H), one of the two minima disappears leading to a saturated ferromagnet (S-FM).

\subsection{The atomic system}
\label{atom}

The magnetic model discussed above can be used to describe the spin sector of an atomic superfluid mixture of two spin states \up and \down.  The correspondence is based on identifying the spin vector components with the population difference $S_3=n_{\uparrow} - n_{\downarrow}$ and the intercomponent coherences with $S_1$ and $S_2$. Given the positive intra- and intercomponent scattering lengths ($a_{\downarrow\downarrow}$, $a_{\uparrow\uparrow}$ and $a_{\downarrow\uparrow}$), we focus on a mixture with $a_{\downarrow\uparrow}^2>a_{\downarrow\downarrow}a_{\uparrow\uparrow}$, which, in the absence of a coupling between the two states, 
undergoes phase separation. The mixture is stabilized by the presence of a coherent radiation (Rabi coupling) with amplitude $\Omega_\text{R}$ and detuning $\delta_B$, which allows for state inter-conversion.  The detuning $\delta_B$ corresponds to the frequency difference between the hyperfine splitting of the two internal levels including the linear Zeeman energy shift, and the frequency of the driving microwave.

\begin{table}[b!]
\centering
   \begin{tabular}{ l | c | c } 
     \hline
     \hline
        Physical Quantity & Magnetic System & Atomic System \\
        \hline
        Anisotropic Interactions & \multicolumn{1}{c|}{$\alpha n$} & \multicolumn{1}{c}{$\kappa n$} \\ 
        Axial field & \multicolumn{1}{c|}{$B_3$} & \multicolumn{1}{c}{$\delta_\text{eff}=\delta_B +  n \Delta$}  \\ 
        \vspace{1mm} && \\
        Transverse  field  & \multicolumn{1}{c|}{$B_1$} & \multicolumn{1}{c}{$\Omega_\text{R}$}\\
        \hline
        \multirow{2}{*}{Spin States} & \multicolumn{1}{{c|}}{$\up$} & \multicolumn{1}{c}{$\ket{2,-2}$}  \\
        & \multicolumn{1}{{c|}}{$\down$} & \multicolumn{1}{c}{$\ket{1,-1}$}\\
        \hline
        Magnetization  & \multicolumn{2}{c}{$\textbf{S}(|\textbf{S}|=n)$ }  \\
        Relative Magnetization  & \multicolumn{2}{c}{$Z=S_3/n$ }  \\
        \hline
        \hline
        \end{tabular}
    \caption{Mapping between magnetic and atomic system.}
    \label{table1}
\end{table}

Table~\ref{table1} illustrates how the Rabi coupling and interaction unbalances map into the components of an effective field in the magnetic model (more details can be found in the Appendix~\ref{app:theory}). The role of the transverse field, $B_1$, is played by $\Omega_\text{R}$. 
The axial component of the external field, $B_3$, has two contributions: the detuning $\delta_B$ and the imbalance of the intra-component atomic interaction energy in the two states $n\Delta\propto (a_{\downarrow\downarrow}-a_{\uparrow\uparrow})n$.  
The difference between intra- and intercomponent scattering lengths $\kappa\propto [(a_{\downarrow\downarrow}+a_{\uparrow\uparrow})/2-a_{\downarrow\uparrow}]$ represents the anisotropic magnetic interactions in the material, uniaxial along direction 3.
Therefore, the resulting effective magnetic field is
\begin{equation}
  \textbf{H}_\textrm{eff}= (  \Omega_\text{R} ,0, \delta_\text{eff} - \kappa  n Z),
\end{equation}
where $\delta_\text{eff}=\delta_B +  n\Delta$.
In the following, we will use this atomic parameter notation to describe the phase diagram.
The precise definition of the parameters $\Delta$ and $\kappa$, which takes into account the geometry of our sample, as well as their experimental estimation, can be found in the Appendices~\ref{app:theory} and \ref{app:calib}.

The crucial role of superfluidity in our experiment is encoded in the term equivalent to the exchange term in \aref{eq:energy} and proportional to $\hbar|\nabla \mathbf{S}|^2/(mn)$ in \aref{eq:extfield}, where \textit{m} is the atomic mass. In the magnetic analogy, this term plays the role of the exchange energy between neighboring spins in a ferromagnet. In our atomic context, it originates from the so-called quantum pressure effect associated to spatial inhomogeneities of the superfluid order parameter and, therefore, appears only in the superfluid state. As a consequence of it, short-wavelength fluctuations of the spin are associated to a sizable increase of the superfluid kinetic energy and are thus inhibited, leading to the observed long-term stability of the hysteretic metastable states. This robustness is to be contrasted to the case of thermal samples, where localized spin rotation would be possible without any appreciable energy increase, making any metastable state prone to fast relaxation.

\section{The experiment}
\label{experiment}

\subsection{Atomic sample}

In contrast to recent works that investigate dynamical properties across a QPT \cite{Nicklas2015,Nicklas2015b} using a rubidium two-component spin mixtures, we realize our two-level system, 
choosing sodium atoms and selecting the hyperfine spin states $\ket{F,m_F}=\ket{2,-2}\equiv\up$ and $\ket{1,-1}\equiv\down$, where $F$ is the total angular momentum and $m_F$ its projection. This yet unexplored spin combination has interesting features for our purposes. 

First of all, such a mixture is stable against spin-changing collisions and possesses intra- and intercomponent scattering lengths ($a_{\downarrow\downarrow}=54.5\,a_0$, $a_{\uparrow\uparrow}=64.3\,a_0$, $a_{\downarrow\uparrow}=64.3\,a_0$, being $a_0$ the Bohr radius) \cite{Tiemann2021}, that make it immiscible in the absence of Rabi coupling.
By combining the chosen spin mixture, sufficiently large peak density $n$, and high magnetic field stability guaranteed by a dedicated magnetic shield \cite{Farolfi19}, we are able to investigate the static properties of the system across the QPT.
In fact, if the typical Zeeman shift associated to the residual magnetic field fluctuations is $\Delta E /\hbar \ll |\kappa|n$, then the ratio between spin interaction energy and the coupling energy  $|\kappa|n/\Omega_\text{R}$ can be finely tuned above and below unity, while keeping the mixture coherent during the whole duration of the measurement.

\subsection{Sample preparation}

We prepare condensates with typical total atom numbers $N=10^6$ and peak densities of $n=10^{14}$ atoms/cm$^3$ in a hybrid trap \cite{Colzi18} inside a magnetic shield that allows for a field stability at the few $\mu$G level \cite{Farolfi19}.
We set an external magnetic field bias of $\SI{1.3}{G}$, necessary to split the magnetic sublevels.
A microwave radiation around 1.769\,GHz is used to coherently couple the two states (\up and \down) with homogeneous and tunable intensity $\Omega_\text{R}$.
The detuning $\delta_B$ is controlled by finely tuning the external field.

\begin{figure}
    \centering
    \includegraphics[width = \columnwidth]{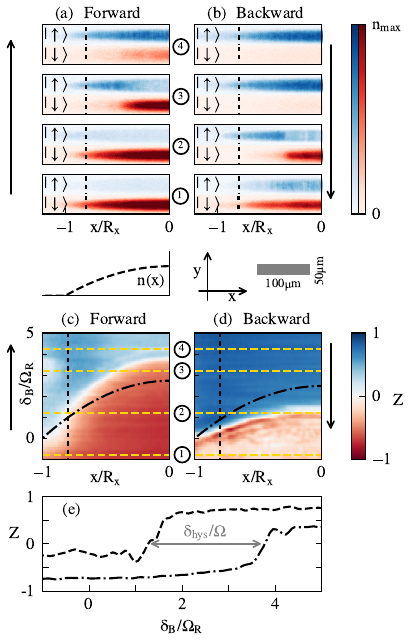}
    \caption{(a-b) Absorption images of the atoms in the states \up and \down (only the left half of the system is shown) for the parameters marked by the yellow lines in Fig.~\ref{fig:fig1}, for forward (a) and backward (b) ramps of $\delta_B$. (c-d) Bare experimental data of the axial magnetization as a function of $\delta_B$ and position $x$ for forward (c) and backward (d) ramps at fixed $\Omega_\text{R}/2\pi = \SI{400}{Hz}$. The solid arrows on the side of the plot indicate the direction of the ramp on $\delta_B$. The yellow dashed lines mark experimental shots shown in panels (a-b), corresponding to number \circledcaption{1}-\circledcaption{4} as in Fig.~\ref{fig:fig1} ($\delta_{B,\circledcaption{1}}/\Omega_\text{R}=-0.8$, $\delta_{B,\circledcaption{2}}/\Omega_\text{R}=+1.2$, $\delta_{B,\circledcaption{3}}/\Omega_\text{R}=+3.2$, $\delta_{B,\circledcaption{4}}/\Omega_\text{R}=+4.2$). The vertical black dashed line marks the position where $|\kappa| n = \Omega_\text{R}$ and the system switches from PM to FM. Dot-dashed black lines in panels (c-d)  mark the local resonance condition  $\delta_B=-n(x)\Delta$. (e) Magnetization of the central 10-$\mu$m-region during a forward (dot-dashed) and backward (dashed) ramp, showing the hysteretic behavior.}
    \label{fig:fig2}
\end{figure}

The degenerate sample is trapped in an elongated optical harmonic trap with trapping frequencies $\omega_{\perp}/2\pi = \SI{2}{kHz}$ and $\omega_x/2\pi = \SI{20}{Hz}$. In this configuration, the BEC is cigar-shaped and presents an inhomogeneous 
axial density profile with a characteristic parabolic shape [see \aref{fig:fig2}(a,b)], typical of a harmonically-trapped system in the Thomas-Fermi (TF) regime, with a longitudinal TF radius $R_x \approx 200$ $\mu$m. 
Our system is fully three dimensional (3D), therefore a mean-field description is justified \cite{Zibold10,Farolfi21,Cominotti2022}. 
However the elongated trapping geometry suppresses the transverse spin excitations and justifies the use of an effective one-dimensional (1D) model. 

Thanks to the smooth density profile $n(x)$ along the longitudinal direction, we can make use of a local density approximation (LDA) for the effective magnetic field $\mathbf{H}_\mathrm{eff}$ through the replacement $n\rightarrow n(x)$. Since the parameters characterizing the phases of the equivalent magnetic system are $|\kappa| n(x)/\Omega_\text{R}$ and $\delta_\text{eff}(x)=\delta_\text{B}+n(x)\Delta$, the spatial dependence of $n(x)$ allows us to observe a spatially resolved phase diagram with different magnetic phases coexisting in the same sample.  The tilted yellow lines in \aref{fig:fig1} represent the regions of the phase diagram which can be experimentally accessed in a single-shot experiment for different choices of $\delta_B$. 

\subsection{Experimental protocol}

In order to experimentally characterize the phase diagram presented in \aref{fig:fig1}, it is important to make sure that the system is always in its local energy minimum.
In all our experiments, we initially prepare the system in a fully polarized state with a large detuning $\delta_B$, and then slowly ramp $\delta_B$ to adiabatically rotate the state to the desired final configuration. Although a very slow rotation would be preferable to maintain adiabaticity in a larger detuning range, especially in the vicinity of the transition, collisions in the mixture reduce the coherence of the sample. The choice of the ramp speed has to be consequently a compromise between adiabaticity and coherence.
In a first set of experiments, we initialize the system in the \down state and linearly ramp $\delta_B$ towards positive values with a constant speed of about 100\,Hz/ms (\textit{forward ramp}). 
In a second set of experiments, a reversed procedure is performed starting from a fully polarized state \up and lowering the value of $\delta_B$  (\textit{backward ramp}). 

For all used $\Omega_R$, the forward ramp starts from an initial detuning of -3.5\,kHz leading to a full ramp time between about \SI{30}{ms} (for $\delta_B/\Omega_R \approx -1$) and \SI{55}{ms} (for $\delta_B/\Omega_R \approx +5$). The backward ramp starts from an initial detuning of +4\,kHz leading to a full ramp time between \SI{20}{ms} (for $\delta_B/\Omega_R \approx +5$) and \SI{45}{ms} (for $\delta_B/\Omega_R \approx -1$). These choices are compatible with a reasonable degree of adiabaticity and with the expected coherence time of about 100\,ms, as estimated from condensate density and collisional properties \cite{Nikuni2003,Harber2002}.

We expect the local magnetization in the low-density tails of the cloud, for which $|\kappa| n(x)<\Omega_\text{R}$, to smoothly change sign as a function of $\delta_B$ (left grey panel of \aref{fig:fig1}), behaving as a PM. On the contrary, the high-density central part of the cloud, for which $|\kappa| n> \Omega_\text{R}$, should remain longer in the initial state during the $\delta_B$ ramp, starting from an initially S-FM configuration, then entering the proper FM phase, and eventually rotating the spin to the other S-FM state once the FM region is over (right grey panel of \aref{fig:fig1}). In the FM region, the presence of a double well allows for the magnetization to have opposite signs depending on the preparation protocols. 
The bifurcation shown in the lower grey panel of \aref{fig:fig1} cannot be observed in a single realization with given $\Omega_\text{R}$ and $\delta_B$, because the condition $\delta_\text{eff}= 0$ is fulfilled only locally.

For each experimental run, information about the spatial spin state is gathered from absorption images of the \up and \down population, with protocols similar to those reported in Ref. \cite{Farolfi21}. Even though the overall condensate fraction of our samples is as low as 30\%, the thermal component is distributed over a larger volume and, thanks to repulsive interactions with the condensate, sits mostly outside this latter~\cite{Goldman1981,Dalfovo99,Mordini2020}.
Through Hartree-Fock calculations, we estimate a condensate fraction in the center of the trap as high as 90\% and we expect that the residual thermal fraction not to play a significant role in the magnetic behaviour because of the low density and short coherence time. In order to extract the properties of the condensate, during the image post-analysis we subtract the thermal component as discussed in detail in Appendix \ref{app:img}. 

Since the tight radial confinement suppresses the transverse spin excitations, as it is clear from absorption images, we focus on the spatial dependence of the relative magnetization $Z$ along the $x$ direction, which is obtained by integrating the magnetization of the two-dimensional (2D) raw pictures along the $y$ direction (a brief discussion on the residual effect due to the transverse dynamics is given in Appendix~\ref{app:2D}).

Examples are shown in \aref{fig:fig2}(a-b). The left-right symmetry of the system leads to the same results on the two sides. While we use both for the statistics, here and in the following, we show only the left part of the cloud to highlight how $Z$ changes in space for increasing $n$ on the horizontal axis.
Labels \circled{1} $\rightarrow$ \circled{4} correspond to four different configurations of the system for increasing $\delta_B$, starting from the system in the ground state \down with large negative $\delta_B$. 
The experiment is repeated also starting from the \up state with large positive $\delta_B$ that is decreased towards negative values (backward, \circled{4} $\rightarrow$  \circled{1}).

\section{Experimental observation of the phase transition}

As a first quantitative measurement, we employ the experimental set-up presented in the previous Section to observe and characterize the quantum phase transition from the PM to the FM state as theoretically presented in Section~\ref{model}. Our study here will address the typical properties of the system stationary state, such as the phase diagram in Section~\ref{diag} and the magnetic response and fluctuation properties in Section~\ref{susce}.

\begin{figure}
    \centering
    \includegraphics[width = \columnwidth]{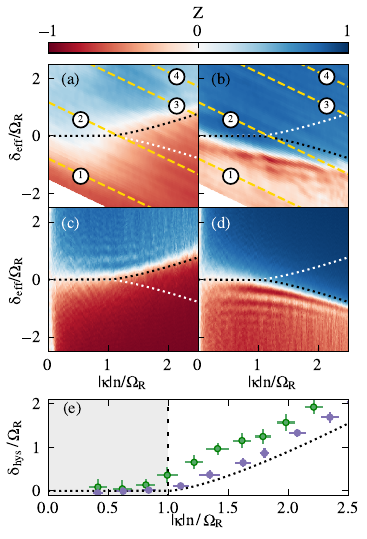}
    \caption{Magnetic hysteresis. (a)-(b) Experimental magnetization data from Fig.~\ref{fig:fig2}(c-d), rescaled according to the \up-\down asymmetry and to the density profile, see main text. White regions in the bottom-left corner are due to a lack of data that manifests when applying the vertical-axis rescaling. (c)-(d) 1D mean-field numerical simulations for the experimental parameters of Fig.~\ref{fig:fig2}(c-d). The dotted black and white lines in panels (c-d) mark the border of the hysteresis region calculated from theory. Yellow dashed lines mark experimental shots shown in panel (a), corresponding to number \circledcaption{1}-\circledcaption{4}, as in Fig.~\ref{fig:fig1}. (e) The width of the hysteresis $\delta_\mathrm{hys}$ is calculated as explained in the Appendix. Green points are experimental data with their uncertainties resulting from the binning procedure and systematic errors. The dotted line stands for theory, while the purple points are results from numerical simulations.}
    \label{fig:fig3}
\end{figure}

\subsection{Phase diagram and hysteresis phenomena}
\label{diag}

Panels (c) and (d) in \aref{fig:fig2} show the experimental axial magnetization as a function of the final applied detuning $\delta_B$ for forward and backward ramps keeping a constant Rabi frequency $\Omega_\text{R}/2 \pi= 400$ Hz. 
The regions where $Z$ changes sign is clearly different for the two protocols. 
The location of the $Z=0$ line along a parabolic-like curve in the ($x,\delta_B$) plane is easily explained: in a fully paramagnetic sample the zero magnetization line would coincide with the locus of points satisfying $\delta_\text{eff}(x)=0$, hereafter referred to as local resonance (indicated by the dash-dotted line in \aref{fig:fig2}); with a harmonic trap in the TF regime, this curve corresponds to a parabola, due to the density-dependent detuning.  In addition to this, the ferromagnetic nature of the cloud shifts the $Z=0$ line from the local resonance and pushes it towards the edges of the hysteresis region, the direction of the shift depending on the sign of the slope of the  $\delta_B$ ramp.

At fixed $\delta_B$, the interface spatially lags behind the local resonance: towards the tail of the cloud (see \circled{3}) for a forward ramp, and towards the center for a backward one (see \circled{2}). 
Thus, along the $x$ direction, the relative magnetization $Z$ zeros between the S-FM region (exterior) and the FM internal region. On the other hand, at fixed $x$, the interface is pushed towards higher values of $\delta_B$, with respect to the local resonance. Figure~\ref{fig:fig2}(e) shows how $Z$ changes around $x\simeq0$, in the case of a forward ramp (dot-dashed) and towards lower values for a backward one (dashed), as it was pictorially represented in the right panel of \aref{fig:fig1}. This behaviour marks the evidence of a hysteresis cycle, observable both as a function of $x$ and $\delta_B$. Since the system undergoes an abrupt discontinuous transition, spin excitations are unavoidable and their presence makes the final magnetization state not reach unitary values.

The raw data in \aref{fig:fig2}(c-d) qualitatively agree with the expected behaviour. 
By plotting the magnetization using the dimensionless quantities, $|\kappa | n(x)/\Omega_\mathrm{R}$ and $ \delta_\text{eff}(x)/\Omega_\mathrm{R}$ for the horizontal and vertical axes, respectively, we obtain the phase diagram reported in Fig.~\ref{fig:fig3}(a)-(b). 

We compare our measurements to a mean-field calculation [panels (c) and (d) of \aref{fig:fig3}] based on two coupled 1D Gross-Pitaevskii equations (GPEs) for the spinor superfluid order parameter $\Psi = ( \psi_\uparrow,\psi_\downarrow)^\top$ (see Appendix~\ref{app:theory} and \ref{app:sim} for more details). Within this formalism, we can properly take into account both the trapping potentials and time sequence used in the experimental protocols. The local spin is given by $\textbf{S} = \text{Tr}(\vec{\sigma}\Psi\otimes\Psi^\dagger)$, with $\vec\sigma$, the Pauli matrices. 
The numerical simulations confirm the observation of a hysteretic region and show a good agreement with the experimental data also for what concerns small structures resulting from the experimental protocol.

\aref{fig:fig3}(e) shows the hysteresis width $\delta_\mathrm{hys}$ (see Appendix~\ref{app:theory} for definition and calculation) as a function of $|\kappa| n/\Omega_\text{R}$, that has been computed analytically [see \aref{eq:hys_width}], numerically [panels (c-d), obtained from simulations performed at 5 different values of $\Omega_\text{R}$] and experimentally (by averaging over more than a thousand shots obtained for different $\Omega_{\text{R}}$). Remarkably, the ultra-stable magnetic environment ensures that the uncertainty on the value of $\delta_B$ is negligible as compared to the relevant parameters of the system, leading to small experimental error on the $\delta_\mathrm{hys}/\Omega_\text{R}$ axis.

The results well capture the presence of hysteresis above the critical point and its monotonic growth for increasing $|\kappa| n/\Omega_\text{R}$.
We checked the role of the transverse directions with 2D GPE simulations (see Appendix~\ref{app:2D}) and found that they explain the residual discrepancy between experiment and 1D GPE simulations or the uniform mean-field theory. 

The discrepancy between the numerical simulation (purple points) and the theoretical expectation (dotted line), attributed to beyond-LDA effects, is reduced by considering slower detuning ramps: if the evolution is not truly adiabatic, spin currents, which are included in the simulations, play a small, although observable, role. 

Finally, it is worth pointing out that the hysteresis phenomena observed in~\cite{Trenkwalder2016} referred to the completely different case of a zero-dimensional single component condensate with attractive interactions in a tunable double-well potential. The crucial novelty introduced by our setup resides on the spontaneous emergence of hysteresis due to strong atom-atom interactions in a spatially extended system, which opens the way to study the interplay of hysteresis with the spatial dynamics. 

\subsection{Magnetic susceptibility and magnetic fluctuations}
\label{susce}

In the vicinity of the phase transition many quantities characterizing the system's response to external parameters diverge. One of these is the magnetic susceptibility $\chi$, which we can extract as the variation of magnetization against variation of $\delta_\text{eff}$ as
\begin{equation}
\chi=\frac{\partial{Z}}{\partial{\delta_\text{eff}}}\bigg\rvert_{\delta_\text{eff}=0}.
\label{chi}
\end{equation}

\begin{figure}[t]
    \centering
 \includegraphics[width = \columnwidth]{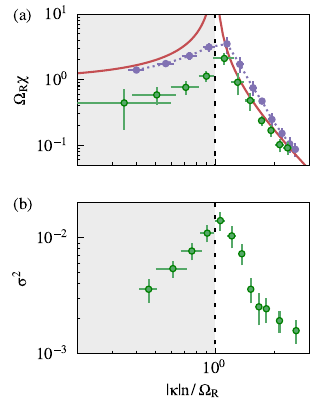}
    \caption{(a) Magnetic susceptibility. Green points are experimental data with their uncertainties resulting from the binning procedure and systematic errors, red line is the theory prediction, purple points connected by dashed line are simulation results. (b) Magnetic fluctuations. The variance of $Z$ is extracted in a central region of the cloud and shows a maximum at $|\kappa| n/\Omega_\text{R} \approx 1$. Error bars are standard variations resulting from averaging different experimental realization and from different binning procedures.}
    \label{fig:fig4}
\end{figure}

Within the universality class of Landau theory, the susceptibility has a finite value at large transverse field where the magnetization follows the applied field, it goes to zero when strong interactions fix the magnetization to \up or \down state, and it diverges at the critical point $|\kappa| n(x)/\Omega_\text{R}=1$, where small variations of the effective field lead to strong changes in $Z$. 

In the homogeneous mean-field approximation (see Eq. \eqref{eq:energy1}), the susceptibility can be written as
\begin{equation}
\label{eq:suscMF}
    \frac 1 \chi =  \left|\frac{\partial\delta_\text{eff}}{\partial Z}\right|_{\delta_\text{eff}=0} = |\kappa| n\begin{cases}
        \cfrac{\Omega_\text{R}}{|\kappa| n} - 1 & |\kappa|n < \Omega_\text{R}, \\
         \left( \cfrac{|\kappa| n}{\Omega_\text{R}} \right)^2-1  & |\kappa|n > \Omega_\text{R},
    \end{cases}
\end{equation}
with the typical asymmetric behaviour of a $\mathbb{Z}_2$ phase transition in the PM and in the FM region \cite{sachdev}. This behaviour is well captured by the experimentally measured $\chi$ (see Appendix~\ref{app:susc}), shown in \aref{fig:fig4} (green dots), where it is compared with the prediction of Eq. \eqref{eq:suscMF} (red lines) and with the numerical solution of noisy GPEs (purple dots), detailed in Appendix~\ref{app:sim}.
To suppress spurious effects arising from inhomogeneity, we restrict the analysis to regions of the sample where the density is nearly constant. 

Both experimental data and simulations do not show a diverging behaviour, but a peak, whose maximum value is slightly shifted on the ferromagnetic side. The absence of a divergence is consistent with finite size effects, while the shift is most probably related to the presence of noise and lack of adiabaticity. Indeed by determining the susceptibility from a GPE without any noise but with dissipation, which kills the fluctuations due to non-adiabatic processes, we still find a peak, but centered at $|\kappa| n/\Omega_\mathrm{R}=1$.

We observe a very good agreement between the numerics and the experimental data on the ferromagnetic side, while the experimentally extracted susceptibility is suppressed on the paramagnetic side, still preserving the right behaviour. We attribute such a discrepancy to the experimental observation that the decoherence is enhanced when the system is nearly unpolarized ($Z=0$), hence affecting the PM side with higher impact. This leads to smaller contrast and could explain the smallness of the extracted value of the susceptibility. 
In the FM region, where instead $Z$ is close to $\pm1$ and the system is observed to be more robust against decoherence, the agreement between the experimental measurement and the model improves.

The measured susceptibility, at least on the FM side, strengthen the previous observation concerning the phase diagram, that the system is well described within mean-field theory\footnote{Since the spin dynamics is mostly along a single direction, one could naively expect the magnetic behaviour to be strongly affected by quantum fluctuations, but such expectation is not supported by the observations.}.
 
The susceptibility was also measured in a spin-orbit coupled BEC from the frequency of the spin-dipole mode across the zero-momentum to plane-wave phase transition \cite{Li2012,Zhang2012}. There, however, the physics was fully dominated by the single particle Hamiltonian and, in particular, did not originate from many-body effects, as witnessed by the fact that the critical point did not depend on the density.

\begin{figure*}
\centering
 \includegraphics[width = \textwidth]{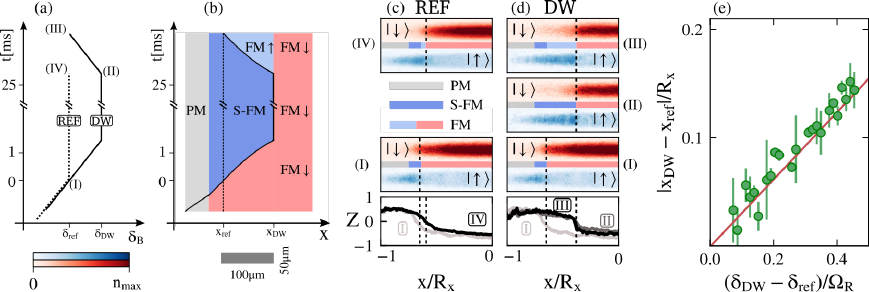}
    \caption{
    Deterministic creation of FM domain walls. (a) Experimental protocol used to create DW through a ramp on $\delta_B$. 
    (b) Schematics of the spatial variation of the phase diagram as a result of the protocol shown in (a). Different regimes are labelled with the same color as in (c) and (d).
    (c) Absorption images of the two components (left half of the system only) at the initial point (I) and after a wait time of 25 ms (IV), when $\delta_B/2\pi=\delta_\text{ref}/2\pi = \SI{1}{kHz}$ [dashed line in panel (a)]. 
    (d) Absorption images of the two components corresponding to the solid line ramp in panel (a), where $\delta_B$ reaches $\delta_{\rm DW}/2\pi =\SI{1.13}{kHz}$ (II) and is then ramped back down to $\delta_\text{ref}$ (III). PM, S-FM and FM regions are illustrated in the line between the absorption images. The third (III) image in panel (c) shows the presence of a DW between two FM domains with opposite magnetization. 
    In both panels (c) and (d) dashed lines marks the position at which $Z=0$.
    (e) Continuous dependence of the position $x_{\rm DW}$ of the DW  with respect to the initial interface position $x_{\rm ref}$ (in units of $R_x$) as a function of ($\delta_{\rm DW}-\delta_{\rm ref}$)/ $\Omega_\text{R}$. The red line is extracted from numerical simulations. Error bars show the experimental uncertainties (horizontal axes) and the shot-to-shot standard deviation (vertical axes).}
    \label{fig:fig5}
\end{figure*}

Another feature of interest in phase transition deals with fluctuations of the order parameter, which, as for the magnetic susceptibility $\chi$, are also expected to diverge at the critical point \cite{sachdev}.
The experimental platform allows us to measure the fluctuations of the relative magnetization $Z$ both in the PM and FM regions. By fixing the detuning $\delta_\text{eff}$ to the local resonance in the central part of the cloud, we measure the variance $\sigma^2$ of $Z$.
We acquire up to 100 realizations for about 20 different values of $\Omega_\text{R}$. As a first step, we calculate $|\kappa| n/\Omega_\text{R}$ for each realization by taking into account the measured atom number in the shot. 
The calculation of magnetic fluctuations is performed by computing the standard deviation of the axial magnetization in a region $w_x = \SI{120}{pixel} \approx\SI{123}{\mu m}$ wide, where the density profile is almost flat, to minimize density-related effects on $\sigma^2$. 
To suppress spurious effects due to the limited resolution of the imaging, we perform the $\sigma^2$ analysis by grouping $N_p$ pixels.
The variance $\sigma^2$ so obtained corresponds to:
\begin{equation}
\label{eq:sigma}
      \sigma^2 = \left\langle \frac{1}{w_x/N_p}\sum^{w_x/N_p}_{i}\left(Z_{i}- \sum^{w_x/N_p}_{j}\frac{Z_{j}}{w_x/N_p} \right)^2 \right\rangle_{N_p},
\end{equation}
where $Z_{i}$ is the relative magnetization of the $i$-th grouping element and $\langle ...\rangle_{N_p}$ is the average over different grouping sizes. The final results plotted in \aref{fig:fig4}(b) are obtained by binning the fluctuation data in a fixed interval of  $|\kappa| n/\Omega_\text{R}$, where the uncertainties are taken as a combination of the standard deviation of the fluctuation between different shots and different binning. They clearly show how the measured variance is maximal at the critical point and reflects the behavior observed for the susceptibility. The same analysis has been performed just outside the condensate in an area containing a thermal atom number comparable to one present in the 120$\times$20-pixel region of interest at the center. The magnetic fluctuations in this thermal component are one order of magnitude smaller ($<10^{-3}$) than those of the condensate, confirming that the fluctuations shown in \aref{fig:fig4}(b) indeed originate from the condensate. Details are given in Appendix\,\ref{app:thermal}.

It is worth noticing that, in general, for a  large homogeneous system, due to the fluctuation-dissipation theorem, the fluctuations of the magnetization of the system and its susceptibility are strictly related. In our system, however, the variation of the number of atoms from shot-to-shot, the finiteness of the system and its inhomogeneity prevent us from a proper quantitative analysis of their relation. 

\section{Deterministic creation of ferromagnetic domain walls}

Another fundamental feature characterizing ferromagnetism is the possibility of forming spatial domains with opposite magnetization. This can take place in a stochastic way via the Kibble-Zurek mechanism during a sudden quench across the PM to FM phase transition \cite{Zurek85,Zurek96,Kibble76,Kibble80}, or by directly engineering the domains with suitable protocols. Different FM domains are separated by domain walls (DW), which constitute low-energy and long-lifetime excitations of the ferromagnet. A review of such investigations in the field of solid-state magnetism can be found in \cite{ALBISETTI2016}. Recently, the spontaneous and deterministic creation of DWs in an effective ferromagnetic BEC under a periodic driving was shown in \cite{yao2022, parker2013}. In these works, however, the ferromagnetic DW was not supported by interactions, but was rather externally created via a spatially varying single-particle potential. 

In our system, we are able to control the size of the FM region of the cloud and, inside it, to deterministically create in a precise yet flexible way DWs where the magnetization $Z$ changes sign, and then control their position at will.
To this purpose, we exploit the dependence of the spatial boundaries of the FM region on the applied detuning $\delta_B$ to control both the position of the DW and the extension of the FM area. Our protocol [\aref{fig:fig5}(a)]  consists in the following steps: 1) ramping  $\delta_{B}$, as for the data in \aref{fig:fig2}, to values $\delta_{\rm DW}$ for which part of the system is in the FM regime; 2) waiting for 25 ms to let the system relax; 3) ramping back to a fixed detuning $\delta_\text{ref}=2.5\,\Omega_\text{R}$ with $\Omega_\text{R}/2\pi=400$\,Hz. Figure~\ref{fig:fig5}(b) shows the position of the boundaries between the different magnetic phases along the longitudinal direction during the $\delta_\mathrm{B}$ ramp.

In Fig.~\ref{fig:fig5}(c) and \aref{fig:fig5}(d), we present the absorption images of the two states for the left half of the sample in case of the two different ramps reported in \aref{fig:fig5}(a). 
As illustrated by the color bar between the absorption images, [which matches the color code used in Fig.~\ref{fig:fig5}(b)], 
the location where the PM region ends remains fixed at the position in the BEC, where $|\kappa|n/\Omega_\text{R}=1$. During the ramp of $\delta_B$, the interface between S-FM and FM moves accordingly, i.e., the magnetic interface moves to higher-density regions (step 1) and then re-enlarges going back to the initial size (step 3). However, during the ramp in step 3, the size of the FM  domain in the \down state remains unchanged,
and a FM domain in the \up state forms in the remaining FM region. 
In this manner, the \up-\down interface, that previously separated S-FM from the FM, becomes a ferromagnetic DW within the FM region, at a position determined by the final detuning $\delta_{\rm DW}$. 

In \aref{fig:fig5}(e), we show the displacement of the DW $x_{\rm DW}$ measured  experimentally from the reference position $x_{\rm ref}$, as a function of $\delta_\textrm{DW}$. The agreement of the experimental results with the 1D GPE numerical simulations is a further indication of the validity of mean-field theory to our atomic system. 

In addition, the smooth and linear dependence observed in this figure demonstrates how the smoothness of the confinement potential allows for the continuous and deterministic control of the DW position via $\delta_{\rm DW}$, without it being pinned by external disorder as it often happens in solids. This key result showcases the promise of our set-up in view of future studies of the quantum relaxation dynamics of the domain wall.

\section{Conclusions and outlook}

In this work, we explored the zero-temperature magnetic phase diagram of a two-component superfluid gas subject to an external coherent Rabi coupling.  
In addition to the critical region, where enhancement of both magnetic susceptibility and fluctuations was detected, a special attention was paid to the ferromagnetic state where metastability and hysteresis features are observed, and domain walls separating different magnetic states are deterministically generated.

The comparison of our results (density profiles, phase diagram, susceptibility) with a zero-temperature mean-field theory seems to indicate that the finite temperature of the superfluid system does not quantitatively affect the behaviour of the QPT, and that the transition is mean-field-like \cite{Roy22}. 
On the one hand, this observation suggests that the spin degrees
are not in thermal equilibrium at the temperature of the gas.  Indeed the sample temperature on the order of 1~$\mu$K is extremely high for the spin sector, in particular much larger then the spin gap. However spin-changing collisions are expected to be very weak, leading to a very large spin collisional time, i.e., long spin relaxation time \cite{Nikuni2003}. This preserves the coherence of the spin sector associated to the initially strongly polarized state. A similar situation for the $\mathbb{Z}_2$-symmetric mixture of Sodium ($|1,\pm1\rangle$) was reported in \cite{Farolfi21,Farolfi21b}.

On the other hand, although the spin physics we are interested in is dominated by the longitudinal dynamics (see Appendix~\ref{app:2D} for the role of transverse dynamics), the system is far from being strictly 1D and has a finite size.  This justifies the good agreement with a mean-field analysis.

In the future, it will be of paramount interest to study the same system in true one and two dimensions, so to unambiguously pinpoint the origin of the mean-field character. If the ferromagnetic transition of our superfluid is in the quantum Ising universality class, a 1D system should show strong deviations from the mean-field results. An important shift in the critical point is expected and the growth of the magnetisation should be very different with respect to the mean-field result, being the critical exponent $\beta=1/2$ in the latter, while $\beta=1/8$ for the 1D quantum Ising model \cite{sachdev}.
It is worth mentioning that, a different spinor superfluid system, with an even more pronounced 1D character than ours, subject to quenches across the transition \cite{Nicklas2015}, showed mean-field critical exponents.
Whether such a result is due to final size effects, closeness to the critical point or finite temperature is still an open question.

Our studies highlight the power of the specific two-component atomic superfluid platform employed here, for a number of key open problems. As a natural first step, one can take advantage of the non conservation of magnetization in the system and the subsequent reinforced quantum fluctuations to analyse their scaling as a function of the subsystem size in the critical region.

Besides the investigation of the static properties of the system in its ground or metastable state immediately after the preparation, the challenge is now to extend the study to the quantum many-body dynamics. 
The superfluid nature of the atomic gas suggests the possibility to investigate magnetism in a novel dissipationless and collisionless regime where the coherence of the two-component superfluid is not affected on the timescale of the experiment by thermal collisions nor by the trap imperfections~\cite{coherencetimecomment, Fava18}. The combination of robust isolation from the environment, and long-lasting quantum coherence in the system will open to explorations of the quantum relaxation dynamics in metastable spinor superfluid.

For instance, in the initial presence of domain walls separating ferromagnetic domains in different states, spin current may develop through the domain wall, so to push the metastable state towards its ground state. The underlying microscopic process may include dissipating the extra energy into the collective excitations of the superfluid, such as spin- or density-phonons \cite{Recati22,Cominotti2022}. 
In the absence of initial ferromagnetic domains, on the other hand, relaxation of the metastable spin superfluid involves, as a preliminary step, a stochastic local spin rotation under the effect of quantum fluctuations and the subsequent spontaneous formation of ground state \textit{bubbles}. The latter should grow, then, according to the previous mechanism, eventually bringing the whole system to its ground state. Beyond its intrinsic interest for quantum statistical mechanics, observing this mechanism will pave the way to the experimental study of false vacuum decay phenomena~\cite{Lagnese21,Marino} and will shine light on processes of crucial cosmological interest~\cite{Georgescu2014,Billam2019,Ng2021}.

\section{Acknowledgements}
We thank S. Giorgini, G. Rastelli, A. Biella, S. Stringari, S. Lannig and M. Oberthaler for fruitful discussions, E. Tiemann for providing us the relevant scattering lengths, and A. Farolfi for experimental contributions in the early stage of the measurements.
We acknowledge funding from Provincia Autonoma di Trento, from INFN through the FISH project, from the Italian MIUR under the PRIN2017 project CEnTraL (Protocol Number 20172H2SC4), from the European Union’s Horizon 2020 research and innovation Programme through the STAQS project of QuantERA II (Grant Agreement No. 101017733), from the European Research Council (ERC) under the European Union’s Horizon 2020 research and innovation programme (grant agreement No 804305), and from PNRR MUR project PE0000023-NQSTI.
This work was supported by Q@TN, the joint lab between University of Trento, FBK - Fondazione Bruno Kessler, INFN - National Institute for Nuclear Physics and CNR - National Research Council.

\appendix

\section{Theoretical framework}
\label{app:theory}

At the mean-field level, a 1D superfluid spin mixture is described by two coupled GPEs for the two order parameters $\psi_\downarrow, \psi_\uparrow$:
\begin{align}
i\hbar \partial_t\psi_\downarrow &= \left(-\frac{\hbar^2\nabla^2 }{2m} + V + g_{\downarrow\downarrow}n_\downarrow + g_{\downarrow\uparrow}n_\uparrow \right)\psi_\downarrow - \frac{\hbar \Omega_\text{R}}{2} \psi_\uparrow\label{eq:gp1}\\
i\hbar \partial_t\psi_\uparrow &= \left(-\frac{\hbar^2 \nabla^2}{2m} + V - \hbar\delta_B(t) + g_{\uparrow\uparrow}n_\uparrow + g_{\downarrow\uparrow}n_\downarrow \right)\psi_\uparrow + \nonumber \\ & - \frac{\hbar \Omega_\text{R}}{2} \psi_\downarrow \label{eq:gp2}
\end{align}
where $m$ is the sodium mass, $g_{\uparrow\uparrow}$, $g_{\downarrow\downarrow}$ and  $g_{\downarrow\uparrow}$ are intra- and inter-component interactions, linked to the s-wave scattering lengths by:
\begin{equation}
    g_{ij} = \frac{4\pi \hbar^2 }{m} a_{ij}.
\end{equation}
Strength and detuning of the coherent coupling are indicated with $\Omega_\text{R}$ and $\delta_B$, while $V$ is the external harmonic potential, trapping the atoms.

In view of analysing the magnetic properties of the mixture, it is convenient to define the spinor $\Psi = (\psi_\uparrow, \psi_\downarrow)^T$ and the density matrix $\rho = \Psi \otimes\Psi^\dagger$. The state of the mixture can be then represented on a Bloch sphere of radius $n = Tr(\rho)$ and encoded in a spin vector
\begin{align}
	\textbf S &= Tr(\vec\sigma \rho ) =
	n \Big( \sqrt{1-Z^2}\cos\phi, \sqrt{1-Z^2}\sin\phi, Z \Big),
	\label{eq:spinvector}
\end{align}
where $\vec\sigma$ is the Pauli matrices vector, $\phi$ the relative phase of the two components and the relative magnetization $Z$ is defined as $nZ = n_\uparrow-n_\downarrow$. The equation governing the dynamics of the spin vector can be derived directly from \aref{eq:spinvector}, \aref{eq:gp1}, \aref{eq:gp2}.
Imposing that the density is uniform and the total (density) current is zero, one obtains \cite{Nikuni2003,Recati22}:
\begin{equation}
	\partial_t \textbf S  = -\textbf H_\textrm{eff}(\textbf S) \times \textbf S .
	\label{eq:spin}
\end{equation}

The state-dependent nonlinear external field
\begin{align}
	\textbf H_\textrm{eff}(\textbf S) &= \big( \Omega_\text{R}, 0,   \delta_B + n \Delta  - \kappa n Z  \big) + \frac{\hbar}{2mn}\nabla^2 \mathbf S 
	\label{eq:extfield}
\end{align}
depends on the interaction constants as:
\begin{align}
    \Delta &\equiv \frac{g_{\downarrow\downarrow}-g_{\uparrow\uparrow}}{2 \hbar} <0  \\
    \kappa &\equiv \frac{g_{\downarrow\downarrow}+g_{\uparrow\uparrow}}{2\hbar}-\frac{g_{\downarrow\uparrow}}{\hbar} <0
\end{align}
and from a density-dependent effective detuning:
\begin{equation}
    \delta_\text{eff} \equiv \delta_B + n  \Delta.
\end{equation}

In the special case of uniform systems, kinetic contributions, which can be ascribed to quantum mechanical currents, can be neglected. Hence, the stationary condition $\textbf H_\text{eff}(\textbf S) \times \textbf S = 0$ translates to
\begin{equation}
    \begin{cases}
        (\delta_\text{eff} - \kappa n Z) \sqrt{1-Z^2} - \Omega_\text{R} Z\cos\phi = 0\\
        \sin\phi = 0
    \end{cases}
    \label{eq:stationarycondition}
\end{equation}
and coincides with the minimization of the energy of the system with respect to both the relative phase and the polarization:
\begin{equation}
     E(Z, \phi) \propto  - \delta_\text{eff} Z + \frac {\kappa n}{2} Z^2 - \Omega_\text{R} \sqrt{1-Z^2}\cos\phi \label{eq:energy2}
\end{equation}
This formula, with the constraint $\phi=0$, is used to calculate the energy profiles shown in the different insets in \aref{fig:fig1}, and the corresponding minimizing polarization in the main graph of \aref{fig:fig1}. The function $E(Z, \phi = 0)$ is symmetric with respect to polarization only at resonance $\delta_\text{eff}=0$. Moreover, it shows a single minimum at $Z=0$ if $|\kappa| n < \Omega_\text{R}$, whereas two degenerate minima at $Z=\pm \sqrt{1-(\Omega_\text{R}/\kappa n)^2}$ in the opposite regime, $|\kappa |n > \Omega_\text{R}$. At the critical point $|\kappa|n=\Omega_\text{R}$, a ferromagnetic QPT takes place, as witnessed by the non-zero value of the polarization, which plays the role of the order parameter for such a transition. 

The criticality of the point $(\delta_\text{eff}, |\kappa|n)=(0,\Omega_\text{R})$ is also confirmed by the divergence of measurable physical quantities, such as the magnetic susceptibility $\chi$, given by \aref{eq:suscMF}.
At finite detuning $\delta_\text{eff}$, the energy profile \aref{eq:energy2} may show two non-degenerate minima: the absolute one describes the ground state of the system, while the local one is associated to a metastable excited state. A hysteresis cycle can therefore be observed by slowly varying the effective detuning from positive to negative values and vice versa. 
The width of the hysteresis region can be computed as follows: according to \aref{eq:stationarycondition}, stationary states are characterized by
\begin{equation}
    \frac{\delta_\text{eff}}{\Omega_\text{R}} = Z\left( \frac{1}{\sqrt{1-Z^2}} - \frac{|\kappa| n}{\Omega_\text{R}}\right) 
    \label{eq:stationarycondition2}
\end{equation} 
In order for the system to be stable, the derivative of this quantity w.r.t. $Z$ (which is inversely proportional to the magnetic susceptibility) must be positive; if the system is ferromagnetic, the derivative of \aref{eq:stationarycondition2} vanishes when the magnetization satisfies $\sqrt{1-Z^2} = (|\kappa| n /\Omega_\text{R})^{-1/3}$. The boundaries of the hysteresis region are therefore given by the detuning values associated to such magnetization:  

\begin{equation}
    \delta_\text{hys} =  2\Omega_\text{R} \left[\left( \cfrac{|\kappa| n}{\Omega_\text{R}} \right)^{2/3}-1\right]^{3/2} .
    \label{eq:hys_width}
\end{equation}

As already mentioned in the main text, the experimental system is confined in a harmonic trap, and consequently has a non-uniform density profile. Since the gas is tightly confined along two out of three spatial directions, it is legitimate to integrate out these two degrees of freedom in order to focus on the dynamics along the longitudinal axis, hereafter indicated as $x$-axis. 
Assuming that $Z$ is only a function of $x$ and integrating Eq.~\eqref{eq:spin} in the $yz$ plane, one obtains a 1D equation, formally identical to Eq. \eqref{eq:spin} with an effective density profile (see, for instance, Ref.~\cite{Farolfi21b}):
\begin{equation}
    n(x) = \frac 2 3 n_0^{3D} \left( 1-\frac{x^2}{R_x^2} \right)
\end{equation}
$n_0^{3D}$ and $R_x$ being the 3D density in the center of the trap and the longitudinal Thomas-Fermi radius, respectively. 

Due to the non-uniformity of the system, the ferromagnetic condition  $|\kappa|n(x) > \Omega_\text{R}$ is only verified at specific real-space positions: there is always a PM region close to the edges of the trap, where the density is smaller. The same holds for the resonant condition $\delta_\text{eff}(x) = \delta_B + n(x)\Delta =0$, due to the $\mathbb Z_2$ symmetry breaking term, $\Delta \ne 0$. More specifically, the locus of resonant points is a parabola in the plane $(\delta_B, x)$, shown as a black dashed line in \aref{fig:fig2}. 

When a detuning ramp is applied to the superfluid mixture, the system does not adiabatically follow the global ground state in the ferromagnetic region, but rather stays in the local metastable minimum until the edge of the hysteresis cycle. In other words, the jump in polarization takes place when $\delta_\text{eff}(x) = \pm \delta_\text{hys}(x)/2$, the sign depending on the direction of the ramp, as shown in \aref{fig:fig2}. 

\section{1D Numerical simulations}
\label{app:sim}

Numerical simulations are performed by exactly solving the 1D GPEs \aref{eq:gp1} and \aref{eq:gp2} in the external harmonic trapping potential. The parameters are chosen to reproduce those of the experiment, taking into account the geometrical renormalization; in particular, $(2/3)|\kappa|n_0^{3D}/h \sim (2/3)|\Delta|n_0^{3D}/h \sim 1.1$ kHz, $\Omega_\text{R}/2\pi = 400$ Hz and $L \sim 200\,\mu$m. The detuning is linearly increased in time at a rate of $100$ Hz/ms. 

The ground state is first found through imaginary-time evolution via the Euler algorithm. We then build the initial state by adding random noise on top of it. The real-time dynamics of this noisy configuration is obtained via a split-step algorithm. We finally average over 100 simulations obtained with different initial noise.
The noise amplitude, which is finally set to $\sim 2\,\%$ of the peak density, is used as a free parameter to best reproduce the experimental data. 

Each simulation produces the magnetization profile in space and time, $Z(x,t)$, which can be straightforwardly interpreted in terms of the local effective detuning: $Z\big(\delta_B(t)+n (x)\Delta  \big)$. The calculation of hysteresis width and magnetic susceptibility is performed as follows: the analysis is applied only to the central half of the cloud $x\in [-R_{x}/2, R_x/2]$, to avoid regions with small density. The magnetization and total density are averaged over $12\,\mu$m-wide windows. Given a window with average density $n^*$, we perform an $arctan$ fit on the function $Z(\delta_\text{eff}^*)$, where $\delta_\text{eff}^* = \delta_B(t)+n^*\Delta $.   
The hysteresis width is given by the shift of the sigmoid center w.r.t. $\delta_\text{eff}^* = 0$, while the susceptibility is found as the sigmoid derivative at $\delta_\text{eff}^*=0$. See the grey panels of \aref{fig:fig1} at $|\alpha| n/B_1 = 0,\, 3$ for illustrative examples of sigmoid functions.

In order to probe both the paramagnetic and ferromagnetic regions, we performed forward and backwards ramps with several values of the Rabi frequency, $\Omega_\text{R}/2\pi = 0.4, 0.6, 0.9, 1.2, 1.5$ kHz, roughly obtaining 130 numerical values for susceptibility and hysteresis width.  
The points appearing in \aref{fig:fig3}(e) and \aref{fig:fig4}(a) are the result of a final binning procedure.

It is worth pointing out that the value of the critical detuning $ n_0^{1D}|\Delta|$ is slightly different for forward and backward ramps, due to the different TF radii for the two components ($g_{\uparrow\uparrow}\neq g_{\downarrow\downarrow}$). In particular, for forward ramps we find $ n_0^{1D}|\Delta|/2\pi \simeq 1.08$ kHz, while, for backward ramps we have $ n_0^{1D}|\Delta|/2\pi \simeq 1$ kHz. This is taken into account in the computation of the effective detuning, that is used to plot \aref{fig:fig3}(c-d), as well as in the calculation of hysteresis width and magnetic susceptibility. 

\section{Effects of transverse directions}
\label{app:2D}

\begin{figure}
    \centering
    \includegraphics[width = \columnwidth]{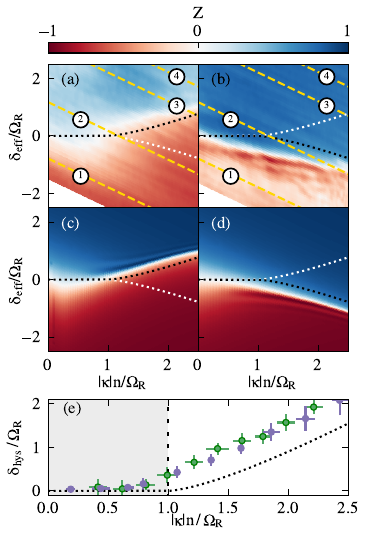}
    \caption{Magnetic hysteresis. As in Fig.~\ref{fig:fig3}, but now the comparison is with numerical simulations including transversal direction. (a)-(b) Experimental magnetization data from Fig.~\ref{fig:fig2}(c-d), rescaled according to the \up-\down asymmetry and to the density profile, see main text. White regions in the bottom-left corner are due to a lack of data that manifests when applying the vertical-axis rescaling. (c)-(d) 1D mean-field numerical simulations for the experimental parameters of Fig.~\ref{fig:fig2}(c-d). The dotted black and white lines in panels (c-d) mark the border of the hysteresis region calculated from theory. Yellow dashed lines mark experimental shots shown in panel (a), corresponding to number \circledcaption{1}-\circledcaption{4}, as in Fig.~\ref{fig:fig1}. (e) The width of the hysteresis $\delta_\mathrm{hys}$ is calculated as explained in the Appendix. Green points are experimental data with their uncertainties resulting from the binning procedure and systematic errors. The dotted line stands for theory, while the purple points are results from numerical simulations.}
    \label{fig:figS1}
\end{figure}

In order to further address the discrepancies with the 1D model discussed in the main text, we decided to performed 2D GPE simulations, which allow us to investigate the role of the transverse direction in the dynamics. 

The numerical code is structured as follows: we first obtain the ground state of the mixture through a conjugate gradient algorithm \cite{antoine2017efficient} and then perform a split-step procedure to determine the real-time evolution of the system. Including fluctuations in a two-dimensional simulation is a computationally expensive task: hence we decided to neglect the effect of noise, which has already been addressed in 1D. Moreover, in order to let the system relax to the ground state after crossing the critical point, we introduced a dissipation term in the simulations: this is realized through a single imaginary-time evolution step of size $\text{d}\tau = \gamma \text{d} t$ every real-time step of size $\text{d} t$, with $\gamma = 0.1$.   

The parameters of the simulation are the same as in the experiment. A comparison between the 2D numerical results and the same experimental data of \aref{fig:fig3}(a-b) is presented in \aref{fig:figS1}.  
The relative 1D magnetization $Z(x)$ has been extracted with the procedure discussed in Appendix~\ref{app:img} for both experimental and simulated data.
The inclusion of an additional dimension is sufficient to obtain quantitative agreement between the numerical and experimental hysteresis width [see panel (e) of \aref{fig:figS1}]. In particular, the critical point appears to be shifted towards $|\kappa| n/\Omega_R < 1$ and the hysteresis region gets larger, as a result of the non-perfect 1D nature of the cloud. 

The much better contrast visible in panels (c,d) of \aref{fig:figS1} with respect to panels (a,b) is due to the presence of losses and absence of decoherence in the 2D numerical simulations. Decoherence does not significantly affect the value of $\delta_\text{hys}$, since the FM region of the cloud is almost always fully polarized in either the \up or \down state. We also verified that the value of $\gamma$ does not modify the relevant properties of the phase transition, but rather has the only effect of damping the spin oscillations excited by the magnetization jump.

\section{Image analysis}
\label{app:img}

\textit{Radial rescaling - } The images of the atomic distributions in the two spin states are taken following a protocol, similar to the one explained in Ref. \cite{Farolfi21}. After a time of flight of 1\,ms, state \up is imaged by standard resonant absorption imaging. Residual atoms in \up are blasted away with a short resonant light pulse. After an additional 1\,ms, \down atoms are transferred to \up by using repumping light and then imaged as before. To calibrate the spin-selective imaging, we ensure that the total atom number remains constant during the transfer between the states. We also radially rescale the image of state \up to match the extension of the image of state \down, by taking into account the extra expansion time. 
The differential expansion time does not introduce relevant changes along the axial direction, thanks to the large trap frequency difference ($\omega_x\ll\omega_\perp$) and the small time of flight, therefore no axial rescaling is needed.\\

\begin{figure}[t!]
    \centering
    \includegraphics[width = \columnwidth]{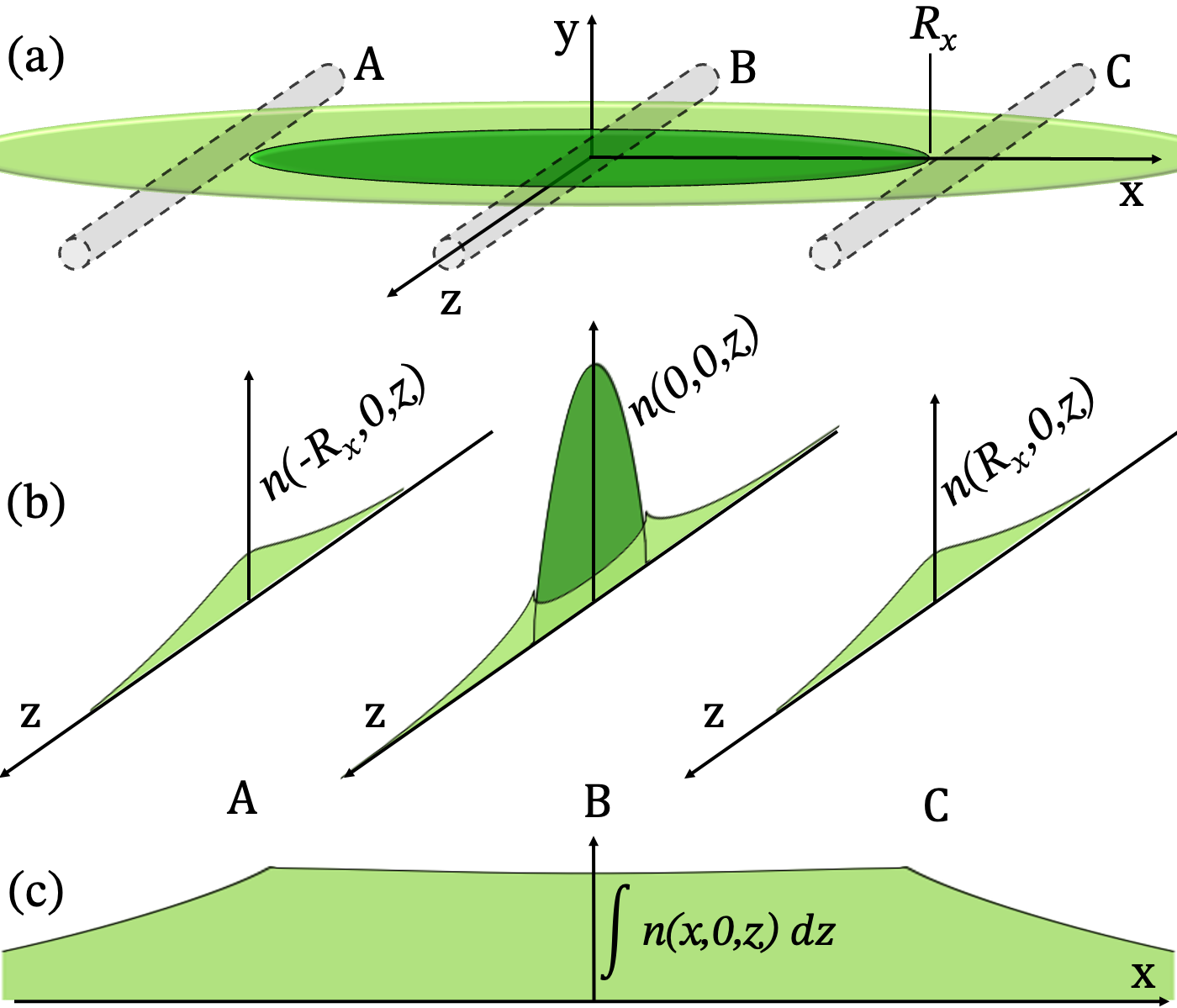}
    \caption{Thermal and condensate atomic distribution. (a) schematic three-dimensional view of the condensate (dark green) and thermal (light green) distribution. A, B and C cylinders highlight the lines of sight of the central atoms and of the thermal atoms just outside $R_x$. 
    (b) Line density profiles along the imaging direction for $x=-R_x,0,+R_x$, calculated using Hartree-Fock theory for our partially condensed (30\%) gas. (c) The integrated density for the thermal component $\int{n(x,0,z)dz}$ has an almost flat distribution in region occupied by the condensate.}
    \label{fig:figS2}
\end{figure}

\textit{Removal of the thermal distribution - } Our 2D images result from the integration of the full atomic density (condensate+thermal) along the line of sight $z$ [see \aref{fig:figS2}(a)]. In order to obtain the densities of the two condensate components, we remove the thermal component in post-analysis. In the following, we assume that the shape of the thermal distribution does not change significantly, while expanding during the small time of flight.

The in-situ atomic distribution of a harmonically trapped bosonic gas is characterized by a dense condensate fraction (dark green) which expels the thermal component (light green) from the center of the trap. Hartree-Fock calculations \cite{Goldman1981,Dalfovo99,Mordini2020} for an ultracold sample with a condensate fraction of 30\% (evaluated against the total trapped atom number) predict a thermal fraction of only 10\% in the central region, as visible in the middle density profile $n(0,0,z)$ shown in \aref{fig:figS2}(b). By numerically integrating only the thermal component density along the line of sight, we obtain an almost flat distribution from one end (A) of the condensate to the other (C) [see \aref{fig:figS2}(c)]. By assuming that the thermal atoms located inside the condensate behave in the same way as the ones located outside, we fit the thermal tails for each state with a Gaussian profile. We remove the contribution of the thermal component 
 from each image by subtracting the fitting Gaussian profile (outside the condensate) and flat top (inside). 
The flat top level is set at a value extracted from the mean value of the thermal component at $(x,y) = (\pm R_x,0)$ and $(x,y) = (0,\pm R_y)$. We do not consider the effects of the asymmetric scattering length on the Hartree-Fock simulations, which would result only in a minor correction \cite{Dalfovo99}.

The obtained condensate distribution is then integrated along $y$. The two populations $n_{\uparrow}$ and $n_{\downarrow}$ are used to calculate the relative magnetization along $x$
\begin{equation}
    Z(x)=\frac{n_{\uparrow}(x)-n_{\downarrow}(x)}{n_{\uparrow}(x)+n_{\downarrow}(x)}.
    \label{eq:Z_x}
\end{equation}

\begin{figure}[t!]
    \centering
    \includegraphics[width = \columnwidth]{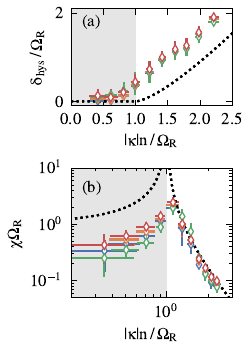}
    \caption{Effects of removing the thermal component. Panel (a) shows the hysteresis width obtained through the four different methods of thermal subtraction, as explained in the text. Correspondingly, panel (b) presents the same comparison performed on the susceptibility. In both panels blue empty symbols correspond to paraboloid, orange to flat top, green to linear and red to gaussian. Both panels show good agreement between the four methods. Black dotted lines are the theory predictions. Error bars are standard variations resulting from averaging different experimental realization and from systematic errors.}
    \label{fig:figS3}
\end{figure}

To verify that the assumptions of a flat top distribution of the thermal atoms in the center of the cloud does not introduce notable effects in the analysis presented in the text, we apply the same methods with four different profiles to remove the contribution of the thermal component: a full Gaussian profile constructed from the Gaussian fit performed on the thermal tails outside the condensate; a linear plane, with a non zero slope along the $x$ direction extracted from the two values of the thermal component at the edge of the BEC, $x = \pm R_x$ (in order to account for spatial asymmetries); the flat top explained just before; an inverse paraboloid, to account for the residual in-site depletion of the BEC.
We extracted the hysteresis width and susceptibility, as done in the main text, for the four different profiles. The results, shown in Fig.\,\ref{fig:figS3}(a,b), indicate that the key ferromagnetic features associated with the phase transition, namely the presence of hysteresis and the divergent susceptibility, are not qualitatively affected by the choice of the thermal profile being subtracted.

\section{Behavior of the thermal component}
\label{app:thermal}
\begin{figure}[t!]
    \centering
    \includegraphics[width = .9\columnwidth]{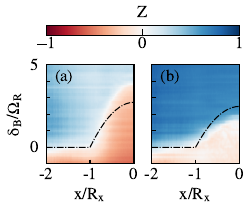}
    \caption{Unprocessed experimental data for forward (a) and backward (b) ramp as in Fig.\,\ref{fig:fig2}(c) and (d) before applying the thermal removal procedure. Dot-dashed black lines mark the local resonance condition  $\delta_B=-n(x)\Delta$.}
    \label{fig:figS4}
\end{figure}

\textit{Magnetization of the thermal component - } Figure \ref{fig:figS4}(a) (forward ramp) and  \aref{fig:figS4}(b) (backward ramp) show the unprocessed magnetization corresponding to the data presented in \aref{fig:fig2}(c-d). Outside $R_x$, the relative magnetization passes from blue to red around zero detuning without showing any $x$ dependence. This corresponds the expectation that the thermal fraction follows the external detuning $\delta_B$ behaving as a gas of non-interacting particles. As expected for a paramagnet, the behaviour does not change between forward and backward ramps and, in particular, does not show any sign of hysteresis. The weaker contrast of the relative magnetization $Z$ seen in \aref{fig:figS4}(a) and (b) is understood as the thermal fraction being more sensitive to decoherence process during the ramp preparation. Note that the coherence time of the thermal fraction is further limited by the additional uncertainty of the effective mean-field spin interaction seen by thermal atoms stemming from the variety of available trajectories through the high density central part of the sample.

\textit{Fluctuations of the thermal contribution - } 
The fluctuations of the thermal component in the central region are not removed by simply subtracting the flat top profile of the average density. 
Their contribution has been analysed based on the spatial distributions in Fig.\,\ref{fig:figS2}(a). 
We select the area corresponding to $A+C$ in the Figure, having a number of thermal atoms equal to the one present in the central $120\times20$ pixel region of interest used in Fig.~\ref{fig:fig4}(b) and evaluate the corresponding fluctuations of the magnetization. The variance  has been then obtained with Eq.~\ref{eq:sigma} and results in a value $\sigma^2<10^
{-3}$ much smaller than the one shown in Fig.~\ref{fig:fig4} for the condensate.

\section{Experimental calibration of \textit{n}}
\label{app:calib}

The determination of $|\kappa| n$ and $n\Delta$ is critical to determine the parameter $|\kappa| n /\Omega_\text{R}$ and, more important, to locate the resonance $\delta_\text{eff}=0$. Due to a fortunate coincidence in collisional parameters of our mixture, $\Delta$ and $\kappa$ differ only at the $10^{-3}$ level. For the two involved hyperfine states, coupled channel calculations provide $a_{11}=54.5\,a_0$, $a_{22}=64.3\,a_0$ and $a_{12}=64.3\,a_0$ \cite{Tiemann2021}. The quantity $n\Delta$ can be experimentally determined either through spectroscopic protocols \cite{Farolfi21} or by locating the resonance position $\delta_\text{eff}=0$ at the center of the cloud in the PM regime, with large $\Omega_\text{R}$, so that hysteresis is absent. We verify the consistency between the two methods and the direct determination of $|\kappa| n$ from the experimentally measured atom number and trap frequencies together with geometrical consideration \cite{Farolfi21b}.

\section{Experimental susceptibility}
\label{app:susc}

In our measurement of $\chi$, we used thousands of experimental scans performed for different values of $\Omega_\text{R}$ with either forward or backward ramps, as the ones presented in \aref{fig:fig2}(c)-(d). To evaluate $\chi$, we make use of the fact that the derivative of the magnetization with respect to $\delta_{\text{eff}}$ is equivalent to the derivative with respect to $\delta_B$. 
\begin{equation}
    \chi=\frac{1}{n}\frac{\partial{s_z}}{\partial{\delta_\text{eff}}}\bigg\rvert_{\delta_\text{eff}=0}
    =\frac{1}{n}\frac{\partial{s_z}}{\partial{\delta_\text{B}}}\bigg\rvert_{\delta_\text{eff}=0}\underbrace{\frac{\partial\delta_B}{\partial{\delta_\text{eff}}}}_{1} \bigg\rvert_{\delta_\text{eff}=0}.
\end{equation}
As a first step, for noise reduction, we spatially average the magnetization $Z$ as well as the total density within a series of 10-pixel-wide windows. For each window, we obtain the value of the magnetization as a function of $\delta_B$ and we perform an $arctan$ plus $linear$ model fit. The estimate for $\chi$ is then extracted as the value of the derivative of the arctan-lin fit at $\delta_\text{eff}=0$. Associated  $|\kappa| n$ is obtained from an averaged density profile of the experimental shot with $\delta_\text{eff}$ closest to zero. 
This procedure results in significant uncertainties for points in the tails of the cloud where the density gradient is large. For this reason, we chose to exclude the outer points from the final binning. 

\bibliography{bibliography.bib}

\begin{thebibliography}{46}
\bibitem [{coh()}]{coherencetimecomment}%
  \BibitemOpen
  \href@noop {} {\bibinfo {title} {{The timescale for the decoherence in the
  spin channel is longer than 100 ms and its is mostly due to the interaction
  with the thermal fraction}}}\BibitemShut {NoStop}%
\end{thebibliography}

\end{document}